\documentclass[sigconf]{acmart}

\usepackage{caption} 
\usepackage{multirow}
\AtBeginDocument{%
  \providecommand\BibTeX{{%
    \normalfont B\kern-0.5em{\scshape i\kern-0.25em b}\kern-0.8em\TeX}}}

\begin{document}

\title{A  Social Search Model for Large Scale Social Networks}


\author{Yunzhong He}
\authornote{Both authors contributed equally to this research.}
\affiliation{%
  \institution{Facebook}
  \streetaddress{1101 Dexter Ave N.}
  \city{Seattle}
  \state{Washtington}
  \postcode{98109}}
\email{yunzhong@fb.com}

\author{Wenyuan Li}
\authornotemark[1]
\affiliation{%
  \institution{University of California, Los Angeles}
  \streetaddress{580 Portola Plaza}
  \city{Los Angeles}
  \state{California}
  \postcode{90095}}
\email{liwenyuan.zju@gmail.com}

\author{Liang-Wei Chen}
\affiliation{%
  \institution{Facebook}
  \streetaddress{1101 Dexter Ave N.}
  \city{Seattle}
  \state{Washtington}
  \postcode{98109}}
\email{benlwchen@fb.com}

\author{Gabriel Forgues}
\affiliation{%
  \institution{Facebook}
  \streetaddress{1101 Dexter Ave N.}
  \city{Seattle}
  \state{Washtington}
  \postcode{98109}}
\email{gforgues@fb.com}

\author{Xunlong Gui}
\affiliation{%
  \institution{Facebook}
  \streetaddress{1101 Dexter Ave N.}
  \city{Seattle}
  \state{Washtington}
  \postcode{98109}}
\email{xunlongg@fb.com}

\author{Sui Liang}
\affiliation{%
  \institution{Facebook}
  \streetaddress{1101 Dexter Ave N.}
  \city{Seattle}
  \state{Washtington}
  \postcode{98109}}
\email{suiliang@fb.com}

\author{Bo Hou}
\affiliation{%
  \institution{Facebook}
  \streetaddress{1101 Dexter Ave N.}
  \city{Seattle}
  \state{Washtington}
  \postcode{98109}}
\email{jsjhoubo@fb.com}

\renewcommand{\shortauthors}{He, et al.}

\begin{abstract}
With the rise of social networks, information on the internet is no longer solely organized by web pages. Rather, content is generated and shared among users and organized around their social relations on social networks. This presents new challenges to information retrieval systems. On a social network search system, the generation of result sets not only needs to consider keyword matches, like a traditional web search engine does, but it also needs to take into account the searcher's social connections and the content's visibility settings. Search ranking should also be able to handle both textual relevance and the rich social interaction signals from the social network. 

In this paper, we present our solution to these two challenges by first introducing a social retrieval mechanism, and then investigate novel deep neural networks for the ranking problem. The retrieval system treats social connections as indexing terms, and generates meaningful results sets by biasing towards close social connections in a constrained optimization fashion. The result set is then ranked by a deep neural network that handles textual and social relevance in a two-tower approach, in which personalization and textual relevance are addressed jointly. 

The retrieval mechanism is deployed on Facebook and is helping billions of users finding postings from their connections efficiently. Based on the postings being retrieved, we evaluate our two-tower neutral network, and examine the importance of personalization and textual signals in the ranking problem.

\end{abstract}



\maketitle

\section{Introduction}
The popularization of social platforms has changed how content is organized on the internet, and has led to the diagram shift from Web 1.0 to Web 2.0 \cite{OReilly2005}. In Web 1.0, users passively consume content from web pages, whereas in Web 2.0, users participate in the creation and sharing of content. Compared to Web 1.0 content, content created on social networks (or Web 2.0) bears more personal aspects. For example, the content may be only visible to a small social group and could require more personal context to understand. User studies also show that compared to web search, users seek information differently on social networks, in which social relation plays a critical role \cite{Evans08,Oeldorf14,Teevan11}.

In classic models of information retrieval systems (such as Web search engines), the main concern is about finding relevant documents based on search keywords. The user dimension is generally a secondary concern and is often introduced into the system at the re-ranking stage based on user profiles \cite{Speretta05,Seig07,Daoud09,Grbovic18}. Although this works well for web pages, it cannot handle the complex social structures of content created on social networks.

Some recent research introduces social relations into information retrieval models by factoring them into user profiles and creating personalized re-ranking models \cite{Carmel09,Vosecky14,Zhao16,Dridi2017} which are often used in microblog search. Among those studies, the social relationship between a user and a document is either used implicitly to construct a better preference-based user profile \cite{Vosecky14,Zhao16,Dridi2017}, or used explicitly as an additive term in the scoring function \cite{Carmel09}. The focus on ranking and microblog search overlooks the fact that for large scale social networks, social relation is an important component to tractably generate the candidate result set to be ranked, due to the private nature of social network content. In addition, treating the social connection as an additive term in ranking offers simplicity but fails to capture the interactions between social signals and textual matches. 

Inspired by its success in artificial intelligence, researchers start using deep learning to process personalization signals in recommendation systems \cite{Wang15,Wide16,He17,Dlrm19} and textual relevance in web search \cite{Dssm13,Dssm-lstm16,Matchtensor17}. Some more recent studies also use deep learning for personalized commerce search \cite{Grbovic18,Haldar19}, which essentially creates neural networks that can jointly handle textual and contextual relevance (e.g. location, interest). Although to our best knowledge, no similar work has been done in the field of social network search, it is natural for us to consider deep learning for the ranking problem due to the rich social contexts behind the results being retrieved. In this paper, we offer the following contributions.
\begin{enumerate}
  \item We introduce a social retrieval mechanism that treats content retrieval as constrained optimization of query rewriting. We use it to help billions of users finding postings from their connections, and we discuss the trade-offs between various factors in this optimization problem using online performance. 
  \item We explore novel neural network architectures that jointly model textual and social relevance for personalized ranking. Based on search results from our production retrieval system and neural network ranker, we examine the importance of different signals in the ranking problem.
\end{enumerate}

\begin{figure}
  \includegraphics[width=8cm]{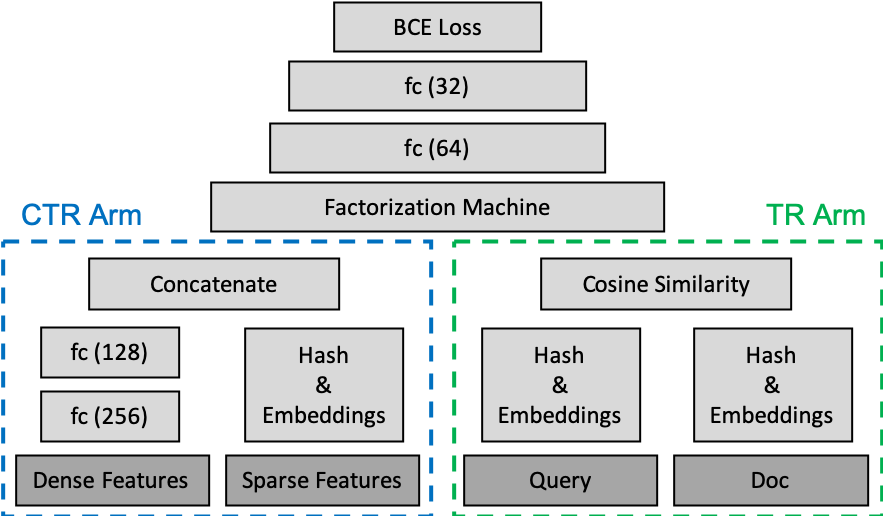}
  \caption{Architecture of our proposed two tower neural network.}
  \label{fig:arch}
\end{figure}

\section{Related Works}

\subsection{Social Search}
Social search refers to the process of finding information based on social connections \cite{survey16}. With the increase in user-generated content on the web, many researchers start to augment traditionally web search engines with social relations or collaborative behaviors for better search experiences \cite{Morris07,Filho10}. Another important area of social search is about searching content on social platforms such as Facebook\footnote{facebook.com} and Twitter\footnote{twitter.com}, which we will refer to as social network search here. In a social network search, social relation plays a more important role than a collaborative web search. Some early works show that having social relations as a term in the ranking objective can significantly improve ranking quality in social network search \cite{Bender08,Carmel09}. In more recent works, social relations are used to construct better user preference models to enhance personalized search ranking in microblog search \cite{Vosecky14,Zhao16,Dridi2017}. The role of the user-author relationship in microblog search ranking is also discussed in \cite{Jiang16,Tabrizi19} with a focus on interests based similarity.

\subsection{Personalized Ranking}
Click-through rate (CTR) prediction is one of the tasks in information retrieval. It focuses on predicting the probability that content would be clicked if shown to the user. Traditionally, people use handcrafted features extracted from Bayesian \cite{richardson2007predicting} or feature selection methods \cite{jahrer2012ensemble,he2014practical}. With the recent developments in deep learning, many CTR prediction methods utilize deep neural networks to reduce the amount of manual feature engineering. For example, neural collaborative filtering \cite{he2017neural} uses a multi-layer perceptron to replace inner product for collaborative filtering. Wide and deep network \cite{cheng2016wide} jointly trains wide linear models and deep neural networks to account for both  memorizations and generalizations. Deep factorization machine \cite{guo17deepfm} utilizes a factorization machine layer so that it can incorporate the power of factorization machines for the recommendation. Deep and cross network \cite{wang2017deep} further replaces the factorization machine layer with cross networks, making it more efficient in learning certain bounded-degree feature interactions.

\subsection{Textual Relevance}
Textual relevance is an extensively studied area in information retrieval which aims to model the semantic similarity between queries and documents \cite{mitra2017neural}. Recent successes of neural methods in the field can be mainly categorized into representation-based and interaction-based methods \cite{guo2019deep}. Early attempts of neural textual relevance models, such as DSSM \cite{huang2013learning} and CDSSM \cite{shen2014learning}, are mainly about learning good representations of the queries and documents, and the similarity measure is based on similarities of the representations. Interaction-based methods, on the other hand, directly model query-document matches at the token level. For example, Arc-II \cite{hu2014convolutional} uses 1-D convolutional layers to model interactions between two phrases. Match-SRNN \cite{wan2016match} introduces the neural tensor layer to model complex interactions between input tokens. MatchPyramid \cite{pang2016text} and PACRR \cite{hui2017pacrr} are inspired by the neural models for the image recognition task - they both view the matching matrix as a 2-D image and use convolutional neural networks to extract hierarchical matching patterns for relevance estimations.


\section{Social Retrieval}
Different from keyword search over web pages, keyword search over a social network requires treating the social dimension as the first-class citizen because the content may be only visible to a small social group on a social network. In addition, social network content is generally more personal and less authoritative, making techniques based on document quality less effective. On a sparse social network like Facebook, without considering social connections in retrieval, the majority of results retrieved may not have any social connections to the searcher and would be either invisible\footnote{Due to user-selected privacy settings, a posting might only be visible to a limited audience. Visibility checks are always enforced on Facebook.}, or irrelevant to the searcher. Even within the domain of all socially connected content, not all social connections are equally relevant. For example, there could be postings from a group that the user joined, but never visited, leading to irrelevant postings retrieved. Given these nuances, the problem of social retrieval is about biasing the retrieval space towards the most relevant connections. Several possible solutions for efficient social network content indexing have been explored in the past \cite{Bjorklund11,unicorn,Bouadjenek13,Chen18}, although their implications on large scale social networks are rarely discussed. In this section, we will use Unicorn \cite{unicorn} as our indexing system, and show how keyword search over Facebook postings can be achieved efficiently with its semantics and a few extended edge types.

\subsection{Problem Definition}
We denote a social network as a directed graph $<U, V>$, where $U$ is the set of nodes, and $V$ is the set of edges. On Facebook, a node can be of different types, such as $person, group, page$, or $posting$. The interactions between nodes are captured by edges. For example, a person can be a friend of another person, which we denote as $<person_1 \xrightarrow{\text{friend-of}} person_2> \in V$. Similarly, a person can make a posting in a group, which we denote as $\{<posting_1 \xrightarrow{\text{authored-by}} person_1>, <posting_1 \xrightarrow{\text{posted-in}} group_1>\} \subseteq V$. Note that we're treating content (posting) and non-content (person, group, page) equally as nodes on a social graph. Since the focus of this paper is on content search, we do want to make the distinction between the two types of nodes. We will refer to person, group and page nodes as entities from now on. Therefore, we have $U = Entities \cup Postings,\ Entities = Persons\ \cup\ Pages\ \cup\ Groups$.
For a searcher $u$,  let us define its social connections as $Conn(u) = \{e\ |\ e \in Entities\ AND\  (<e, u> \in V\  OR\ <u, e> \in V\ OR\ u=e)\}$. And we can now formally define its socially connected postings as $ConnPostings(u) = \{p\ |\ p \in Postings\ AND\  \exists\ e \in Conn(u)\ s.t.\ (<e, p> \in V\  OR\ <p, e> \in V)\}$. Here, $ConnPostings(u)$ represents all the postings on a social network that are directly connected to entities that are within 1-degree of connections between the searcher. As discussed above, we may not want to consider all social connections as the result sets could be noisy, so we would like to filter down to the postings from a set of good social connections $GoodConn(u) \subseteq Conn(u)$, leading to the final retrieval space $GoodConnPostings(u) = \{p\ |\ p \in Postings\ AND\  \exists\ e \in GoodConn(u)\ $ $s.t.\ (<e, p> \in V\  OR\ <p, e> \in V)\}$. Thus, the goal of social retrieval is to retrieve postings with keyword matches like traditional web search does, but restricted to the retrieval space of $GoodConnPostings(u)$ by enforcing a bias at query rewriting stage towards $GoodConn(u)$.

\subsection{Unicorn}
We use Unicorn \cite{unicorn} as our indexing system. In Unicorn, an edge on a social network is indexed as a prefixed term in a document's inverted index. For example, edge $<person_1 \xrightarrow{\text{friend-of}} person_2>$ is indexed as {\fontfamily{qpl}\selectfont friend:2} in the inverted index of $person_1$ \footnote{We denote entity an unique identifier $x$ with  $entitiy_x$ }. To support keyword search on postings, we extended Unicorn with edge types as shown in Table \ref{table:edges}, which captures various relationships a posting can have with a person, a group, and a page. As for text terms in a posting, they are indexed the same way as traditional web search engines do. For example, a posting about "Billie Eilish" will have the following indexing terms {\fontfamily{qpl}\selectfont text:billie, text:eilish}. \footnote{Extensive studies have been done on optimizing the indexing policies for text terms, but we would like to keep it simple here as it is not the focus of this paper.}

During retrieval, we use the standard query language provided by Unicorn - s-expressions that support logical operators such as $AND/OR$. Suppose searcher $person_0$ has best friends $person_1$ and $person_2$, is an active member of $group_3$, and maintains a Facebook page $page_4$, then a possible set of good social connections could be $GoodConn(u)=\{0, 1,2,3,4\}$. Thus the condition of enforcing posting from $GoodConn(u)$ can be expressed as the following query string.
\begin{quotation}
 {\fontfamily{qpl}\selectfont
 (or \\
 \phantom{} \hspace{0.5cm} involves:0 \\
 \phantom{} \hspace{0.5cm} authored-by:1 \\
 \phantom{} \hspace{0.5cm} authored-by:2 \\
 \phantom{} \hspace{0.5cm} group-of:3 \\
 \phantom{} \hspace{0.5cm} page-of:4 \\
 )
 }
 \end{quotation}
 
\begin{table}[]
\begin{tabular}{|l|l|}
\hline
\textbf{Edge-Type}   & \textbf{Description}                            \\ \hline
authored-by & A posting is authored by a person      \\ \hline
involves    & A posting involves a person in any way \\ \hline
page-of     & A posting is posted in a page          \\ \hline
group-of    & A posting is posted in a group         \\ \hline
\end{tabular}
\caption{Extended edge types}
\label{table:edges}
\vspace*{-\baselineskip}
\end{table}

\subsection{Social Query Rewriting}
We have now formulated the problem of social retrieval as finding an optimal query string to enforce postings from $GoodConn(n)$. Because this is independent of the query string of enforcing text term matches, we will refer to this process as "social query rewriting". More specifically, for a searcher $u$, and a query {\fontfamily{qpl}\selectfont query-keyword-match}, we enforce the social conditions by rewriting it into a final query string of the form {\fontfamily{qpl}\selectfont (and query-keyword-match query-social-match(u))}. For example, after social query rewriting, the query string for keyword "Billie Eilish" will be the following
\begin{quotation}
 {\fontfamily{qpl}\selectfont
 (and \\
 \phantom{} \hspace{0.5cm} (or \\
 \phantom{} \hspace{1cm} text:billie \\
 \phantom{} \hspace{1cm} text:eilish \\
 \phantom{} \hspace{0.5cm} ) \\
 \phantom{} \hspace{0.5cm} (or \\
 \phantom{} \hspace{1cm} involves:0 \\
 \phantom{} \hspace{1cm} authored-by:1 \\
 \phantom{} \hspace{1cm} authored-by:2 \\
 \phantom{} \hspace{1cm} group-of:3 \\
 \phantom{} \hspace{1cm} page-of:4 \\
 \phantom{} \hspace{0.5cm} ) \\
 )
 }
 \end{quotation}
Now that we have formulated the problem of social retrieval as an optimal query rewriting problem, the next thing is to define the optimality condition and provide an optimization mechanism. Since we are optimizing for a candidate generation process, we want to focus more on recall rather than precision. However, higher recall generally requires a larger retrieval space to look at, leading to higher computational cost on the index servers. So finding an optimal query rewriting is essentially a constrained optimization that maximizes recall under a fixed budget of CPU cost.

\subsection{Ground Truth Recall Set} 
Before diving into the final optimization formulation, we need to define a measure of the recall that we are trying to optimize. We sampled a ground truth set of 2k ideal results of the form $<query, searcher,$ $ideal\ result>$ through rater data. The ideal result does not necessarily come from Search Engine Results Page (SERP), as raters may be searching a piece of content seen on a friend's profile. This way we can get rid of the presentation bias. We anonymized and featurized the dataset into the form $feature(searcher, author)$, which contains no user-identifiable information, but rather indicators of the social relationship between the user and the author of the ideal result. Table \ref{table:social_features} illustrates the features we used. When a search request is issued, the information is pre-fetched as the request's meta-data from Facebook's graph database \cite{tao}, which makes them readily available in real-time. For model training, we also sampled non-ideal results from their corresponding SERPs as negatives.

\subsection{Constrained Optimization for Optimal Social Rewriting}
Given the small size of training data and feature space, we use a linear model to optimize the connection terms. For each social prefix $p$, we want to solve for the optimal social feature weights $w^*_p$, and a threshold $t^*_p$ that maximizes the recall in the ground truth set $D_G$ under a fixed CPU cost $k$. More formally,
\begin{equation*}
\begin{aligned}
& \underset{w_p, t_p}{\text{maximize}}
& & recall_{D_{G}}(expr(w_p, t_p)) \\
& \text{subject to}
& & cpu\_cost(expr(w_p, t_p)) < k
\end{aligned}
\end{equation*}
in which $expr(w_p, t_p)$ is the s-expression parameterized by feature weights $w_p$ of a linear model that selects the best connection terms, and a bound on the number of social connections $t_p$.

To solve for $t_p^*$, we assume an uniform retrieval space incurred by each social prefix (i.e. each friend, group and page can provide as many results matching the text constraints as all other friends, groups and pages do, respectively), which makes the CPU cost from each prefix effectively a function of the number of connections with the prefix. This assumption is an oversimplification as some groups may contain more postings than others. However, we found $t_p$ to be the dominating factor of CPU cost when it is aggregated over all search sessions. Since we budget CPU cost at an aggregated level, this is in fact a valid assumption. And to maximize the recall, we simply ran parameter sweep experiments online and obtained the maximal $t_p$ under budget $k$. 

As we fix $t_p$, this constrained optimization becomes a simple maximization of $recall_{{Conn(D_{G}})}@t_p^*(LM(w_p))$. Namely, we want to maximize the top $t_p^*$ recall of the connections that a posting comes from in the ground truth set $D_G$, when the connections are ranked by a linear model parametrized by weights $w_p$. We then solve $w_p^*$ using linear regression.

Essentially, what this constrained optimization does is approximate the optimal connection space $GoodConn(u)$, using a ground truth dataset sampled from $GoodConnPostings(u)$. And it is done through first fixing its size, and then selecting its most likely members. The weights and thresholds are computed offline, but the inference on optimal connections is performed online because of the dynamic nature of social connections.

\begin{table}[]
\begin{tabular}{|l|}
\hline
\textbf{Feature}                           \\ \hline
Whether this user/page/group is recently visited                             \\ \hline
Time since last visit of this user/page/group \\ \hline
Whether this is a liked page                          \\ \hline
Whether this is a joined group \\ \hline
Social network coefficient between the searcher and author \\ \hline
\end{tabular}
\caption{Social features for query rewriting}
\label{table:social_features}
\end{table}

\subsection{Results}
We deployed this rewriting system on Facebook search, and it is now powering the retrieval of postings from a searcher's connections. The retrieval system had evolved from naively including postings from all friends and groups, to a heuristics-based approach (i.e. order by interaction recency) that also enables postings from followed pages, and to the constrained optimization approach that allows more general connection types such as liked or recently visited pages, and achieved good quality improvement with even CPU cost savings. Table \ref{table:rewrite_results} shows the relative improvements from its post-launch backtesting compared against other approaches, where we rank connections based on the recency in the searcher's last interaction as the baseline. We report the overall CTR on postings from friend and group/page. Note that for this backtesting we fix $t_p$ for a fair evaluation of social rewriting quality, and all of the results were ranked by our production ranking model before being displayed to the searchers.

From Table \ref{table:rewrite_results} we can see that CPU cost does not differ significantly between different methods, confirming our hypothesis that CPU cost is a function of $t_p$. And our linear model indeed outperforms any heuristics-based methods like interaction recency or social graph coefficient. We believe that being able to aggregate different real-time interactions from a social network is an advantage of this online linear model, and is where its gain comes from. We have learned from this backtesting and other past experiments that, in general, strong real-time signals like recent interactions can outperform more sophisticated offline models, while the best performance was often achieved when the two were combined.


\begin{table}[]
\begin{tabular}{|l|c|c|c|}
\hline
\textbf{Method} & \textbf{CPU} & \textbf{CTR(friend)} & \textbf{CTR(group/page)}    \\ \hline
Recency & - & - &  -            \\ \hline
Social Coef & (p>0.05) & 0.32\%\ (p ={1.2e-4}) & 0.94\%\ (p ={6.8e-5}) \\ \hline
Linear Model & (p>0.05) & \textbf{0.56\%\ (p ={1.2e-6})} & \textbf{2.07\%\ (p ={3.3e-5})}              \\ \hline
\end{tabular}
\caption{Post-launch performance of social query rewriting}
\label{table:rewrite_results}
\vspace*{-\baselineskip}
\end{table}

\section{Two-Tower Neural Network for Personalized Ranking}

After a set of socially connected postings are retrieved, they need to be ordered based on their relevance for the best user experience. Since content on a social network is highly personal, social relevance should be factored into the ranking objective as much as textual relevance. Although social relevance was considered in the ranker from section 3.5, a key difference here is that the aforementioned ranker only ranks the connections to retrieve from but cannot compare the relevance between two postings to display. Therefore, we want to explore result-level ranking models that can jointly handle social and textual relevance here.

\subsection{Multi-stage Ranking}
Because result-level ranking can have non-trivial CPU cost, we use a multi-stage ranking in our production system. Our stage 1 ranker is a gradient-boosting decision tree (GBDT) model that uses standard textual relevance features like BM25 \cite{bm25}, as well as social relevance features. And the stage 2 ranker is a neural network that leverages more sparse features on top of the stage 1 features, using a DLRM-like \cite{Dlrm19} architecture. We observed a big search quality improvement when we launched the stage 2 neural network ranker, even without the additional sparse features. For the sake of space, we will focus our discussions on the architecture explorations for the neural network ranker here. And the exploration is based on our production neural network architecture.

\subsection{Motivation}
While extensive studies have been done on neural IR models, the focus has been on learning better text representations and query-document interactions \cite{huang2013learning,shen2014learning,hui2017pacrr,mitra2017neural,guo2019deep}. In the settings of personalized web search, personalization signals generally come from historical search behaviors, and the focus is usually on preference-based user profiling \cite{Speretta05,Seig07,Daoud09,Dridi2017}. In social network search, however, personalization signals are so much richer and contain an extra dimension of social relations. This motivates us to look at the click-through rate (CTR) models from recommendation systems, as the scope of personalization is closer to CTR prediction problems rather than traditional web search ranking. However, textual relevance does still play an important role in ranking, so naturally, to combine the best of the two, we proposed a two-tower approach which contains a CTR model trained from personalization features, and a textural relevance model trained from n-grams of query and document tokens. In addition, we decided to train the model with click data, not only because of its availability but also because any 3rd-party labeled data cannot capture the social context of the original searcher.

\subsection{Two-Tower Architecture}
The overall architecture of our two-tower network can be expressed as follows.
\begin{center}
$\Phi(q,d,u) = g(NN_{tr}(\psi(q), \eta(d)), NN_{ctr}(f_{d}(u, d), f_{s}(u, d)))$
\end{center}
where $NN_{tr}$ is a typical query-doc textural relevance (TR) model, modeling the semantic similarity between a query $q$, and a document $d$, based on their embedding representation $\psi(q)$ and $\eta(d)$. $NN_{ctr}$, on the other hand, is a CTR model capturing the contextual relevance between a user $u$ and the document $d$ using a hand-tuned user-doc dense feature vector, $f_{d}(u,d)$, and sparse feature vector $f_{s}(u,d)$. The output of $NN_{tr}$ and $NN_{ctr}$ are both vectors, which we denote as $x_{tr}$ and $x_{ctr}$. Note that $x_{tr}$ is a vector because each dimension of $x_{tr}$ represents the cosine similarity between one type of query n-gram (e.g. query bi-gram, query tri-gram) and one type of document n-gram (e.g. doc-title-trigram, doc-body-bigram). Finally, $x_{tr}$ and $x_{xtr}$ are concatenated into vector $x = x_{tr} \cdot x_{ctr}$ as the input to a final matrix factorization layer, to represent feature-feature interactions. So we have
\begin{center}
$\Phi(q,d,u) = g(x_{tr}, x_{ctr}) = w^Tx + x^Tupper(VV^T)x$
\end{center}
Note that different from standalone TR or CTR models, there isn't any softmax layer in $NN_{tr}$ or $NN_{ctr}$ to normalize the final vector representation into a probability value. Instead, we use the vector representation directly into a matrix factorization layer.
Although by design the two-tower architecture could support any TR and CTR models, in practice we use DLRM \cite{Dlrm19} as our CTR tower, and a simple cosine similarity model as our TR tower, as illustrated in Figure \ref{fig:arch}.
The neural network is trained with click data, which is a collection of pairs $<q, u, d, y>$, in which $y \in \{1, 0\}$ representing a user click ($y=1$), or no click ($y=0$). Therefore the neural network can be trained by minimizing a BCE loss just like a neural factorization machine \cite{guo17deepfm}.

\subsection{Representation Details}
To learn the query representation $\psi(q)$ and document representation $\eta(d)$, we extract n-grams from their raw texts, and feed them into randomly initialized embedding lookup matrices. Note that $\psi(q)$ and $\eta(d)$ don't share the same embeddings, as we believe queries and documents have different textual characteristics.

To encode signals outside of textual relevance, we use a feature vector $f_{d}(u,d)$ with around 30 dimensions in our experiments. $f_{d}(u,d)$ is essentially a concatenation of many hand-tuned features we designed, and those features can be categorized as user-side features, doc-side features, and user-doc features. We also use two user-side sparse features for better personalization, encoded as sparse vectors in $f_{s}(u,d)$. Table \ref{table:features} illustrates those features.
\begin{table}[]
\begin{tabular}{|l|l|}
\hline
\textbf{Feature}                                    & \textbf{Type}      \\ \hline
If the post is authored by the searcher    & user-doc  \\ \hline
If the post comes from a friend            & user-doc  \\ \hline
If the post comes from a joined group      & user-doc  \\ \hline
If the post comes from a followed page     & user-doc  \\ \hline
If the searcher has recently seen the post & user-doc  \\ \hline
Global click count from SERP of a post     & doc-side  \\ \hline
If the post contains a photo               & doc-side  \\ \hline
Age of a post                              & doc-side  \\ \hline
Number of comments on a post               & doc-side  \\ \hline
Number of friends of the searcher          & user-side \\ \hline
Number of followees of the searcher        & user-side \\ \hline
Number of SERP impressions of the searcher                        & user-side \\ \hline
Region id of the searcher                       & user-side, sparse \\ \hline
City id of the searcher                       & user-side, sparse \\ \hline
\end{tabular}
\caption{Hand-crafted features used in the CTR arm}
\label{table:features}
\vspace*{-\baselineskip}
\end{table}

\section{Experiments}
\subsection{Dataset}
Given the personal nature of social content search, it is difficult to create any labeled dataset to evaluate our models. So instead, we use clicks sampled from real search traffic for model training and evaluation. When a search query is issued, for the results shown, their n-gram features from raw texts, as well as the feature vectors $f_{d}(u,d)$ and $f_{s}(u,d)$, are computed online. They are then sampled and logged along with any click activities on them as our training and evaluation data. Any user identifiable information is omitted in the final dataset. So the dataset is a collection of <session-id, hashed n-grams, $f_{d}(u,d)$, $f_{s}(u,d)$>, where session-id is a unique identifier of the search session, which can contain many documents displayed. We sampled 120 millions of data from a 30-day window for model training. To mimic the production behavior, the evaluation set is sampled from some dates after the 30-day window of training data, containing 6 million rows in total, as we find empirically that evaluating on future data yields a closer estimation of a model's online performance.

\subsection{Hand-tuned Textual Relevance Features}
To evaluate how our models can learn textual relevance end to end from raw texts, we adopt several popular query-doc relevance features from information retrieval for comparison. Table \ref{table:trfeatures} shows the textual relevance features we used. For simplicity we will refer to features in Table \ref{table:features}, features in Table \ref{table:trfeatures}, and raw text n-grams as CTR features, TR features, and n-gram features respectively.

\begin{table}[]
\begin{tabular}{|l|}
\hline
\textbf{Feature}                           \\ \hline
BM25                              \\ \hline
AVG TF-IDF                            \\ \hline
Position of the last matched term \\ \hline
\end{tabular}
\caption{Textual Relevance Features}
\label{table:trfeatures}
\vspace*{-\baselineskip}
\end{table}

\subsection{Experiment Setup}
To investigate how social/personalization signal and textual relevance signal can influence the final search ranking results, we train and evaluate our model under the following settings: (1) DLRM with CTR features only; (2) DLRM with TR features only; (3)  DLRM with CTR + TR features; (4) cosine similarity with n-gram features only; (5) two-tower with CTR + n-gram features (6) two-tower with CTR + TR + n-gram features.

Note that under our two-tower design, the CTR arm is essentially a DLRM model without a final softmax layer, so experiment (1) (2) and (3) are essentially the $NN_{ctr}$ tower adding a softmax layer to generate a probability value for ranking. On the other hand, the textual relevance arm adopts a DSSM structure with two heads to process query and document respectively. Setting (4) uses textual relevance arm only. Similarly, we also add a softmax layer for ranking in (4). Setting (5) and (6) use two-tower structures to process social/personalization and textual relevance signals. The only difference is that in setting (6), we also use TR features in the CTR arm,  so that it has the richest signals. A detailed model architecture of our proposed method can be found in Table \ref{table:arcdetail}.

Our model is implemented in Pytorch and trained on 2 Tesla M40 GPU. We use an Adam optimizer with the initial learning rate of $0.01$. We also use batch normalization to ease the training, and add a dropout layer after each fully connected layer for regularization.

\begin{table*}[h]
\begin{tabular}{|l|c|l|c|l|}
\hline
                                & \multicolumn{2}{c|}{\textbf{CTR Arm}}                                                                                                          & \multicolumn{2}{c|}{\textbf{Textual Relevance Arm}}           \\ \hline
\textbf{Input}                  & \multicolumn{1}{l|}{Dense Features}                                                                                        & Sparse Features   & Query                                  & Document          \\ \hline
\multirow{3}{*}{\textbf{Model}} & \begin{tabular}[c]{@{}c@{}}2-Layer MLP:\\ fc(256), batchnorm, relu\\ dropout (0.2)\\ fc(128), batchnorm, relu\end{tabular} & Hash \& Embedding & \multicolumn{1}{l|}{Hash \& Embedding} & Hash \& Embedding \\ \cline{2-5} 
                                & \multicolumn{2}{c|}{Concatenate}                                                                                                               & \multicolumn{2}{c|}{Cosine Similarity}                     \\ \cline{2-5} 
                                & \multicolumn{4}{c|}{\begin{tabular}[c]{@{}c@{}}Factorization Machine\\ 3-Layer MLP:\\ fc(64), batchnorm, relu\\ dropout (0.2)\\ fc(32), batchnorm, relu\\ fc(1)\end{tabular}}                               \\ \hline
\textbf{Output}                 & \multicolumn{4}{c|}{Logistic Layer (click probability)}                                                                                                                                                     \\ \hline
\end{tabular}
\caption{Detailed model architecture.}
\label{table:arcdetail}
\end{table*}

\subsection{Evaluation}

We evaluate our model performance on 0.1 million search results seen by the users, where clicked results are labeled as positives, and non-clicks as negatives. We use ROC-AUC and Normalized Discounted Cumulative Gain (NDCG) as the evaluation metrics here. See \textbf{\autoref{table:evaluation}} for the detailed numbers.

\textbf{CTR and TR features:}
We first focus on evaluating the effectiveness of CTR and TR features in the context of search ranking. We evaluate the features using a DLRM model, which is effectively our two-tower model without the TR arm. We compare the model performance using social features, TR features, and CTR+TR features. The model with the lowest AUC/NDCG uses only TR features, which indicates that manually designed textual relevance features are not expressive enough for a social search problem. Adding CTR and features further improves the AUC from 59.01\% to 62.16\% and NDCG from 80.66\% to 81.55\%.  Overall, we observe that personalization signals are important for the social search ranking model. 
 
\textbf{N-gram Features:}
Next, we experiment with text n-gram features using a simple DSSM-like model, in which each n-gram will have its embeddings.  Using n-gram features alone outperforms TR features by at $1.7\%$ AUC, likely because embeddings improve the expressiveness of the representations. For example, for query "kitten" and a document titled "cat videos", traditional TR features may consider it as a total mismatch. However, such similarity could be modeled in the space of text embeddings.

\textbf{Two-tower architecture}: Finally, we test our proposed two-tower method using CTR + n-gram features. Our proposed model achieves $64.49\%$ AUC, which is another $2.0\%+$ ROC-AUC increase over any of the single tower approaches (with the highest NDCG). That is, combining social and textual relevance signals with embeddings yields the best performance. In addition, combining the n-gram features with TR features further improves AUC from 64.49\% to 65.02\%. 

Overall, we conclude that social/personalization and textual relevance signals are both important for post ranking, and manually engineered textual relevance features are not as effective as text embeddings learned from n-grams. Our proposed two-tower architecture, which jointly models personalization and textual relevance, yields the best ranking based on ROC-AUC and NDCG of click prediction data.

\begin{table}[] 
\begin{tabular}{|l|c|c|}
\hline
\textbf{Inputs}   & \multicolumn{1}{l|}{\textbf{ROC-AUC}}  &\textbf{NDCG}\\ \hline
CTR dense features      & 61.74\%     &       81.30\%          \\
TR dense features       & 59.01\%        &      80.66\%         \\
CTR + TR          & 62.16\%          &    81.55\%       \\ \hline
n-gram sparse features  & 60.81\%          &   80.80\%           \\ \hline
CTR  + n-gram      & \textbf{64.49\%}            &     \textbf{81.67\%}       \\
CTR  + TR  + n-gram & \textbf{65.02\%}           &  \textbf{81.98\%}               \\ \hline
\end{tabular}
\caption{Model performance under different settings.}
\label{table:evaluation}
\vspace*{-\baselineskip}
\end{table}

\section{Conclusion}
In this paper, we present the challenges of social search on large scale social networks such as Facebook. We argue that it is important to tractably generate a meaningful result set is to bias retrieval towards a set of good connections, and we solve this through a constrained optimization problem of query rewriting. We also demonstrate the effectiveness of this approach by comparing online performance across different methods.

Finally, we introduce a neural learning-to-rank model, which jointly learns personalization and textual relevance, in the social network search domain. We argue that the scope of personalization signals in a social network search is similar to recommendation systems, while textual relevance still plays an important role as it is in classic information retrieval models. So we propose a two-tower neural network that combines the best of the two, and our experiments show that it indeed outperforms state-of-the-art models like DLRM, with substantial gains coming from our two-tower design.

\bibliographystyle{ACM-Reference-Format}
\bibliography{references}


\end{document}